\documentclass[aps, prb, reprint]{revtex4-2}

\pdfoutput=1 
\usepackage{graphicx}
\usepackage{color,amsmath}
\usepackage[normalem]{ulem}
\usepackage[utf8]{inputenc}

\usepackage[colorlinks,linkcolor=blue,citecolor=blue,urlcolor=blue]{hyperref}
\usepackage{siunitx}
\usepackage{comment}
\usepackage{layouts}
\usepackage[acronym]{glossaries}
\glsdisablehyper

\begin{document}
\title{Observation of quantum Hall interferometer phase jumps due to changing quasiparticle number}

\author{Marc~P.~Röösli}
\email{marcro@phys.ethz.ch}
\affiliation{Solid State Physics Laboratory, Department of Physics, ETH Zurich, 8093 Zurich, Switzerland}

\author{Lars~Brem}
\affiliation{Solid State Physics Laboratory, Department of Physics, ETH Zurich, 8093 Zurich, Switzerland}

\author{Benedikt~Kratochwil}
\affiliation{Solid State Physics Laboratory, Department of Physics, ETH Zurich, 8093 Zurich, Switzerland}

\author{Giorgio~Nicolí}
\affiliation{Solid State Physics Laboratory, Department of Physics, ETH Zurich, 8093 Zurich, Switzerland}

\author{Beat~A.~Braem}
\affiliation{Solid State Physics Laboratory, Department of Physics, ETH Zurich, 8093 Zurich, Switzerland}

\author{Szymon~Hennel}
\affiliation{Solid State Physics Laboratory, Department of Physics, ETH Zurich, 8093 Zurich, Switzerland}

\author{Peter~Märki}
\affiliation{Solid State Physics Laboratory, Department of Physics, ETH Zurich, 8093 Zurich, Switzerland}

\author{Matthias~Berl}
\affiliation{Solid State Physics Laboratory, Department of Physics, ETH Zurich, 8093 Zurich, Switzerland}

\author{Christian~Reichl}
\affiliation{Solid State Physics Laboratory, Department of Physics, ETH Zurich, 8093 Zurich, Switzerland}

\author{Bernd~Rosenow}
\affiliation{Institute for Theoretical Physics, Leipzig University, Leipzig D-04009, Germany}

\author{Werner~Wegscheider}
\affiliation{Solid State Physics Laboratory, Department of Physics, ETH Zurich, 8093 Zurich, Switzerland}

\author{Klaus~Ensslin}
\affiliation{Solid State Physics Laboratory, Department of Physics, ETH Zurich, 8093 Zurich, Switzerland}

\author{Thomas~Ihn}
\affiliation{Solid State Physics Laboratory, Department of Physics, ETH Zurich, 8093 Zurich, Switzerland}

\date{\today}

\newacronym[longplural={quantum dots}]{QD}{QD}{quantum dot}
\newacronym[longplural={quantum point contacts}]{QPC}{QPC}{quantum point contact}
\newacronym[longplural={charge stability diagrams}]{CSD}{CSD}{charge stability diagram}
\newacronym[longplural={two dimensional electron gases}]{2DEG}{2DEG}{two dimensional electron gas}

\begin{abstract}
We measure the magneto-conductance through a micron-sized quantum dot hosting about 500 electrons in the quantum Hall regime. In the Coulomb blockade, when the island is weakly coupled to source and drain contacts, edge reconstruction at filling factors between one and two in the dot leads to the formation of two compressible regions tunnel coupled via an incompressible region of filling factor $\nu=1$. We interpret the resulting conductance pattern in terms of a phase diagram of stable charge in the two compressible regions.
Increasing the coupling of the dot to source and drain, we realize a Fabry-P\'{e}rot quantum Hall interferometer, which shows an interference pattern strikingly similar to the phase diagram in the Coulomb blockade regime. We interpret this experimental finding using an empirical model adapted from the Coulomb blockaded to the interferometer case. The model allows us to relate the observed abrupt jumps of the Fabry-P\'{e}rot interferometer phase to a change in the number of bulk quasiparticles. This opens up an avenue for the investigation of phase shifts due to (fractional) charge redistributions in future experiments on similar devices.
\end{abstract}

\maketitle

\section{Introduction}
Quantum dots with few hundreds of electrons operated in the quantum Hall regime are a laboratory of intricate quantum Hall physics. This is particularly evident at filling factors $\nu$ between one and two, where two energetically separated spin-polarized Landau levels exist. Ground-breaking experiments \cite{brown_resonant_1989,mceuen_transport_1991,mceuen_self-consistent_1992,mceuen_coulomb_1993} and theory \cite{Dempsey1993} have established the self-consistent reconstruction of the electron distribution inside the quantum dots into compressible and incompressible regions as a result of Landau-level quantization and Coulomb interactions. Theoretical work \cite{evans_coulomb_1993} supported this interpretation and suggested a phase-diagram of stable charge in the parameter plane of the plunger-gate and the magnetic field. Further experimental work found that the magnetic field tunes the tunnel coupling across incompressible regions \cite{van_der_vaart_time-resolved_1994,van_der_vaart_time-resolved_1997}. Increasing the magnetic field reduces the tunnel coupling to such an extent that time-resolved tunneling between the inner compressible region of the upper Landau level and the outer compressible region of the lower Landau level can be detected on time scales of seconds \cite{van_der_vaart_time-resolved_1997}. Later measurements detected and verified the predicted phase-diagram \cite{heinzel_periodic_1994,fuhrer_transport_2001,chen_transport_2009,liu_electrochemical_2018}.

Complementary interest in this system arose in the context of interferometry. Coupling the quantum dot strongly to the leads allowed experimentalists to operate it as a Fabry--P\'{e}rot quantum Hall interferometer in the integer quantum Hall regime \cite{van_wees_observation_1989,taylor_aharonov-bohm_1992,bird_precise_1996}. Some experiments found integer fractions of Aharonov--Bohm periods \cite{godfrey_aharonov-bohm-like_2007,camino_aharonov-bohm_2005-1}, which the theory in Ref.~\onlinecite{rosenow_influence_2007} describes. Later experiments report the transition from Coulomb-dominated to Aharonov--Bohm dominated interference \cite{zhang_distinct_2009,ofek_role_2010,Sivan2016} when the device size is taken to very large values.
Experiments in the fractional quantum Hall regime promise opportunities to measure the fractional statistics of anyonic quasiparticles \cite{camino_aharonov-bohm_2005,zhou_flux-period_2006,camino_quantum_2007,camino_$e/3$_2007, lin_electron_2009, ofek_role_2010, mcclure_fabry-perot_2012, nakamura_aharonovbohm_2019, willett_measurement_2009, willett_alternation_2010, willett_interference_2019}.

In this paper, we investigate a quantum dot at filling factors $1\lesssim\nu\lesssim 2$ that can be tuned continuously from the Coulomb blockade regime into the interferometer regime. In the Coulomb blockade, only the outermost compressible region, belonging to the Landau level lowest in energy, couples to source and drain, and leads to observable conductance resonances. The electron-by-electron depopulation of the upper Landau level leads to intermittent shifts of the observed resonances. From the resulting pattern, we construct and model the phase diagram of stable charge configurations. We extract the edge excitation energy from temperature-dependent measurements and find it to be significantly larger than the estimated single-particle level spacing at zero magnetic field. Reducing the filling factor from two to one exhibits signatures of the reduction of tunneling through the incompressible $\nu=1$ region. This  allows us to operate the dot in a `quenched' regime with a fixed number of electrons in the upper L andau level, such that the Coulomb blockade pattern of the outer compressible region can be observed in isolation. Taking the system gradually to the interferometer regime, the signatures of the phase diagram remain visible, indicating the close relation between the physics in the two regimes of operation \cite{Stern2010} despite the disappearing charge quantization in the outer compressible region. Conductance resonances turn into Fabry--P\'{e}rot resonances of the interferometer, and the depopulation of the upper Landau level leads to characteristic phase jumps in the interference pattern \cite{Sivan2016}. The observed value of the phase jump is larger than $\pi$, in agreement with the observation of Coulomb dominated interference patterns \cite{halperin_theory_2011}. Tuning the interferometer to the quenched regime and thus eliminating phase jumps allows us to measure the signatures of Aharonov-Bohm interference in an otherwise Coulomb dominated regime. Adapting the model for the weakly coupled dot to the interferometer case leads to a consistent description of the observed behavior.
Such detailed observation and modeling of interference phase shifts would allow to disentangle interaction induced and anyonic phase shifts in future fractional quantum Hall devices even in the Coulomb dominated regime.

\section{Experimental Setup}
We use an AlGaAs/GaAs heterostructure hosting a \gls{2DEG} $\SI{130}{nm}$ below the surface etched into the shape of a Hall bar. The \gls{2DEG} is contacted by annealed AuGeNi ohmic contacts. It has the electron mobility $\mu = \SI{5.3e6}{cm^{2}/Vs}$ at the temperature $T=\SI{25}{mK}$ and the electron density $n = \SI{1.3e11}{cm^{-2}}$.
The electron density can be changed by applying a voltage to an overgrown pre-patterened back gate $\SI{1}{\micro m}$ below the \gls{2DEG} that extends under the whole Hall bar \cite{Berl_Structured_2016}.

\begin{figure}[htbp]
\includegraphics[width=\columnwidth]{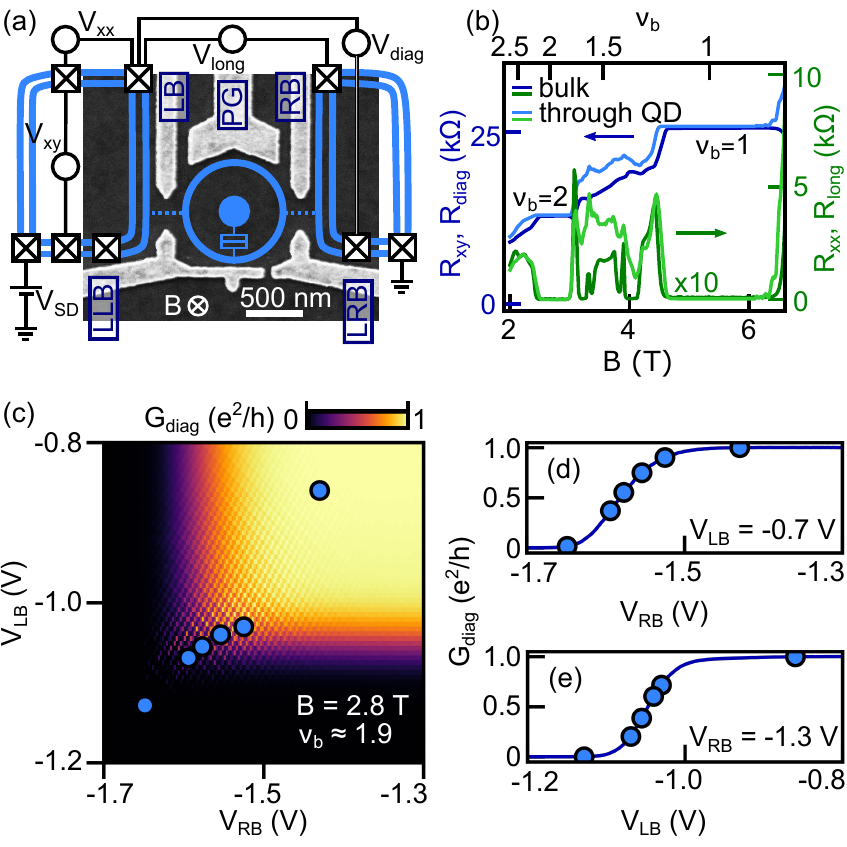}
\caption{(a) Scanning electron micrograph of the \acrfull{QD} device before evaporation of the second top gate layer. The top gates appear in light gray, the uncovered AlGaAs/GaAs heterostructure in dark gray. A magnetic field $B$ is applied perpendicular to the sample surface. The added schematic shows compressible edge regions (blue) for a filling factor $\nu_\mathrm{b} \approx 2$, the arrangement of ohmic contacts, the applied source--drain voltage $V_\mathrm{SD}$, and the measured voltages $V_{xx}$, $V_{xy}$ (bulk), and $V_\mathrm{long}$, $V_\mathrm{diag}$ (across the \gls{QD}). In addition, we measure the current $I_\mathrm{SD}$. (b) Longitudinal and transverse resistances $R_{xx}=V_{xx}/I_\mathrm{SD}$ and $R_{xy}=V_{xy}/I_\mathrm{SD}$ across the bulk (dark blue and dark green) with all gate voltages at zero, compared to $R_\mathrm{long}=V_\mathrm{long}/I_\mathrm{SD}$ and $R_\mathrm{diag}=V_\mathrm{diag}/I_\mathrm{SD}$ measured with all gates energized except RB and LB. (c) Conductance $G_{\mathrm{diag}}=I_\mathrm{SD}/V_\mathrm{diag}$  as a function of $V_{\mathrm{LB}}$ and $V_{\mathrm{RB}}$ at $B = 2.8~\mathrm{T}$ ($\nu_\mathrm{b} \approx 1.9$). Blue points indicate gate voltage settings, where later measurements in Fig.~\ref{fig7} are taken. (d) and (e): Line-cuts of (c) with conductance $G_{\mathrm{diag}}$ vs. $V_\mathrm{LB}$ or $V_\mathrm{RB}$, with the respective other barrier gate transmitting the quantized conductance $e^2/h$.} 
\label{fig1}
\end{figure}

The inner structure of the device is shown in Fig.~\ref{fig1}(a). We lithographically define metallic top gates (light grey) on the surface of the heterostructure (dark grey). By applying negative voltages to the top gates, we deplete the electron gas underneath and form a \gls{QD}. Depletion occurs around $\SI{-0.4}{V}$. We place an additional top gate covering the area of the \gls{QD} in a second gate layer insulated by a \SI{15}{nm} thick $\mathrm{Al_2O_3}$ oxide layer from the other top gates. This additional top gate as well as the back gate are kept grounded for all the measurements shown in this paper.

The presence of the back and the additional top gate influences the electrostatics of the device, leading to a flatter potential in the quantum dot interior, and sharper confinement at its edges. In addition, they tend to reduce the electron-electron interaction by screening. In previous experiments, gating potentials and capacitances of similar samples were optimized in such a way \cite{choi_robust_2015, sivan_interaction-induced_2018, nakamura_aharonovbohm_2019, willett_interference_2019}.

Our sample is measured in a wet dilution refrigerator at the electronic temperature $T_\mathrm{e} = \SI{25}{mK}$.
All measurements presented are performed on the same sample. However, the results were reproduced on another sample fabricated on a different wafer. The sample presented here was measured in three different cooldowns. While the general behavior of the device was the same in all cooldowns, the gate voltage values needed to reproduce a specific regime varied between the three cooldowns.

\section{System characterization}

In a first characterization step, we demonstrate the presence of quantum Hall states in the two-dimensional electron gas, in particular in the region of the \gls{QD}, and quantify the density reduction arising from the application of confining gate voltages to the top gates.
In Fig.~\ref{fig1}(b), we show the longitudinal and transverse resistances $R_{xx}$ and $R_{xy}$ of the electron gas measured outside the \gls{QD}-region as a function of a magnetic field $B$ applied perpendicular to the sample plane [see schematic in Fig.~\ref{fig1}~(a)]. In $R_{xy}$ (dark blue) we observe quantized Hall plateaus for integer bulk filling factors $\nu_{\mathrm{b}} = 1$ and $2$ at magnetic fields, where $R_{xx}$ (dark green) vanishes.
We also observe fractional quantum Hall states at $\nu_{\mathrm{b}} = $4/3, 5/3  between $\nu_{\mathrm{b}} = 1$ and $2$.
Additionally, we measure the diagonal and longitudinal resistances $R_\mathrm{diag}$ and $R_\mathrm{long}$ (shown in light blue and green) through the \gls{QD} region with voltages $V_\mathrm{RB}=V_\mathrm{LB}=0$ (open barriers), $V_\mathrm{LLB}=V_\mathrm{LRB}=\SI{-0.75}{V}$, and $V_\mathrm{PG}=\SI{-0.5}{V}$. This voltage setting allows us to probe quantum Hall states in the strongly confined spatial region, where the quantum dot will later be tuned into the Coulomb blockade or the interferometer regime. The diagonal resistance $R_\mathrm{diag}$ exhibits quantum Hall plateaus at integer filling factors slightly shifted to lower magnetic fields compared to the bulk curve, corresponding to a few-percent reduction of the density in the region where the \gls{QD} is formed due to the voltages applied to the top gates. We do not observe clear fractional states in the transport through the \gls{QD} region.

Simulations of the zero magnetic field electron density and electrostatic potential distribution in the device (not shown) are performed using COMSOL. They are based on a numerical solution of Poisson's equation in three dimensions and treat the screening properties of the electron gas within the Thomas-Fermi approximation. The simulations confirm the experimentally detected reduction of the electron density in the quantum dot region by a few percent as compared to the bulk, given the gate voltages applied in the experiment. They also predict that the peak density in the quantum dot region is reduced by at most 10\% compared to the bulk, when negative voltages are applied to the gates LB and RB in order to tune the \gls{QD}-region into the Coulomb blockade regime. Throughout the paper, we always quote the bulk filling factor $\nu_{\mathrm{b}}$ when describing the measurements, because the exact filling factor in the \gls{QD} is experimentally unaccessible. The values of $\nu_{\mathrm{b}}$ will overestimate the true filling factor in the \gls{QD} by $5-10\%$. The simulations further indicate that the \gls{QD} is populated with 400--650 electrons for all measurements shown in this paper.

In the next step, we demonstrate how we control and quantify the tunnel coupling between the quantum dot region and the bulk electron gas. To this end, we keep the gate voltages on the gates LLB, LRB, and PG as before, and control the coupling of the \gls{QD} to the source and drain regions by changing the voltages applied to the barrier gates LB and RB [see Fig.~\ref{fig1}(a)]. Figure~\ref{fig1}(c) shows the conductance $G_\mathrm{diag}=1/R_\mathrm{diag}$ of the \gls{QD} as a function of $V_\mathrm{LB}$ and $V_\mathrm{RB}$ at $B = \SI{2.8}{T}$ corresponding to the bulk filling factor $\nu_{\mathrm{b}}=1.9$. The bright region (upper right) has conductance $e^2/h$ indicating that both barriers fully transmit the $\nu = 1$ edge channel. Reducing the voltage on either of the barrier gates pinches off the respective barrier and results in a vanishing \gls{QD} conductance. The \gls{QD} is in the Coulomb blockade regime when both barriers are close to pinch-off and exhibit a transmission well below one. In contrast, the system is a quantum Hall interferometer, if the transmission of both barriers and thereby the transmission of the whole device is close to one, resulting in a total conductance close to $e^2/h$.
The measurements shown later in Fig.~\ref{fig7} are performed with the barrier gate voltages marked by the blue points, which span the range from the Coulomb blockaded to the interferometer regime.

At a given gate voltage setting $(V_\mathrm{LB},V_\mathrm{RB})$ we estimate the transmission of each individual barrier using the line-cuts of Fig.~\ref{fig1}(c) shown in (d) and (e) of the same figure. The two cuts are taken at extreme voltages, where the transmission of one barrier is one, while the other barrier transmission is tuned from zero to one. The transmission of the right (left) barrier at a specific voltage $V_\mathrm{RB}$ ($V_\mathrm{LB}$) is read from Fig.~\ref{fig1}(d) [Fig.~\ref{fig1}(e)] as the conductance value at this voltage divided by the conductance quantum $e^2/h$. This procedure assumes that the cross-talk between the gate LB (RB) and the right (left) barrier transmission is weak. This assumption is justified based on the almost perfectly horizontal (vertical) orientation of lines of constant $G_\mathrm{diag}$ in Fig.~\ref{fig1}(c), best discernible at $G_\mathrm{diag}\approx e^2/2h$.

\section{Coulomb blockade regime}
\begin{figure*}[t]
\includegraphics[width=\textwidth]{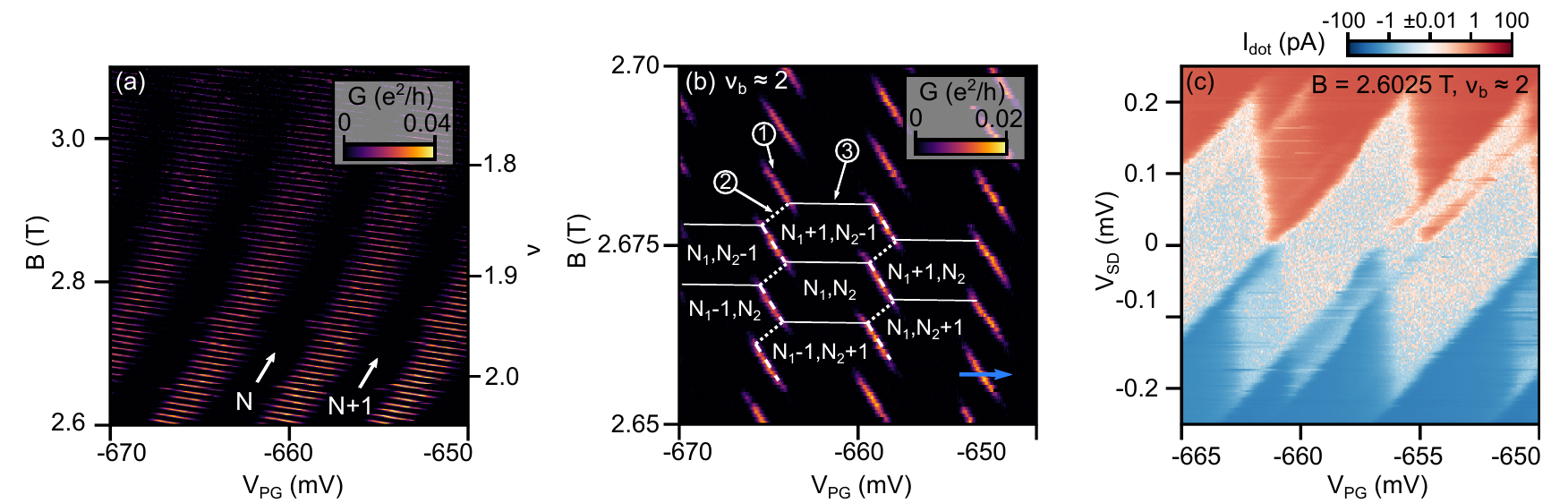}
\caption{(a) Conductance of the \gls{QD} as a function of the plunger gate voltage and the perpendicular magnetic field. (b) Close-up of (a) for a smaller range in magnetic field interpreted as a charge stability diagram (CSD). The lines separate regions of constant charge $(N_1, N_2)$ where $N_1$ and $N_2$ correspond to the electron number in the outer and inner compressible region respectively. Three different transitions are indicated. (c) Coulomb diamond at $\nu_{\mathrm{b}} \approx 2$ showing the current through the \gls{QD} on a logarithmic scale (sign-conserving, cut off at $\pm\SI{10}{\femto A}$) as a function of the plunger a gate voltage and the applied bias voltage.} 
\label{fig2}
\end{figure*} 

We now study the quantum dot coupled weakly to source and drain with transmissions much smaller than one.
Figure~\ref{fig2}(a) shows the two-terminal conductance $G=I_\mathrm{SD}/V_\mathrm{SD}$ of the \gls{QD} as a function of the plunger gate voltage and the magnetic field around a filling factor $\nu \lesssim 2$. In this regime, the contact resistances are negligible compared to the device resistance and the two-terminal measurement gives an accurate conductance of the \gls{QD}. Bands of conductance resonances are observed that shift to more positive gate voltages as the magnetic field increases. They are decorated by finer conductance resonances with a slightly negative slope that are tuned mainly by the magnetic field. Vertical cuts at constant $V_\mathrm{PG}$ would give bunches of closely spaced conductance resonances as a function of magnetic field, separated by larger regions of zero current, as reported in Ref.~\onlinecite{van_der_vaart_charging_1994}. Horizontal cuts at constant magnetic field lead to closely spaced double peaks (see again \cite{van_der_vaart_charging_1994}).

The zoom into a small magnetic field range is shown in Fig.~\ref{fig2}(b). On this scale, the observed conductance resonances form a highly periodic pattern similar to those reported in Ref.~\onlinecite{fuhrer_transport_2001} in a similar device at filling factors $2<\nu<4$, in Ref.~\onlinecite{chen_transport_2009} for $\nu<3$, where a single-electron transistor was employed as a charge detector, and in Ref.~\onlinecite{Sivan2016} at filling factors $1<\nu<2$. 

Coulomb blockade diamond measurements at constant magnetic field [see Fig.~\ref{fig2}(c)] confirm, that single-electron charging of the quantum dot is responsible for the observed resonances. The charging energy of the quantum dot of about \SI{200}{\micro eV} is plausible for the given dot size. The plunger-gate period of about \SI{6}{meV} per electron is in agreement with our zero magnetic field COMSOL simulations. The plunger gate lever arm extracted from the experiment is $\alpha_\mathrm{PG}=0.036$.

Our interpretation of the data in Fig.~\ref{fig2} follows the insights of the seminal Ref.~\onlinecite{mceuen_self-consistent_1992}, which showed experimentally that the commonly used constant-interaction model for a quantum dot cannot account for the observations at $\nu<2$.
The charge distribution inside the \gls{QD} is influenced by the occupation of two spin-split Landau levels. Their bare Zeeman-splitting may be enhanced by exchange effects \cite{Nicholas88}. In an attempt to closely mimic the charge distribution at $B = 0$ in the quantum dot, which minimizes the electrostatic energy, the quantum dot reconstructs self-consistently into compressible and incompressible regions. In particular, the Landau level higher in energy tends to form a compressible region, separated by an incompressible region of filling factor $\nu=1$ from a compressible ring formed by states of the lower energy Landau level [see Fig.~\ref{fig1}(a)]. This reconstruction was theoretically described within a capacitance model by Evans, Glazman and Shklosvskii \cite{evans_coulomb_1993}. They suggested a phase diagram with hexagon-shaped regions of stable integer charge (in units of $e$) in the two Landau levels, describing the one-by-one redistribution of single electrons from the upper to the lower Landau level with increasing magnetic field, and the addition of single electrons to the dot with increasing gate voltage.
This phase diagram was later successfully used for the qualitative interpretation of experimental data \cite{heinzel_periodic_1994,chen_transport_2009,baer_cyclic_2013,liu_electrochemical_2018}.

Like these authors, we assume that the incompressible $\nu=1$ region in the quantum dot has a quantized Hall conductance of $e^2/h$, whereas the Corbino conductance between the inner compressible region of the upper Landau level and the outer compressible region of the lower Landau level is small compared to $e^2/h$, such that the electron numbers in these regions are integer numbers. The observed conductance resonances in Fig.~\ref{fig2} arise from single-electron tunneling between the leads and the outer ring-shaped compressible region of the quantum dot. Direct tunneling between the lead and the inner compressible region is not observed, because the presence of the incompressible $\nu=1$ region reduces the rate of these processes below observable values. However, single electrons can redistribute (tunnel) between the inner and outer compressible regions on sufficiently large time scales, which are still smaller than the time-scale of the measurement in Fig.~\ref{fig2}, in order to establish thermodynamic equilibrium within the dot. Within the picture of spin-split Landau levels, such a tunneling event requires the reversal of the spin of the tunneling electron, which reduces the tunnel rate below that of hypothetical spinless electrons. Since tunneling through a potential barrier usually conserves the spin of the tunneling electron, further ingredients such as spin-orbit interaction \cite{golovach08,hofmann17} or hyperfine interaction \cite{wald94,hanson_spins_2007} are required to enable spin-flip tunneling processes. The net result of the charge transfer from the inner to the outer compressible region with increasing magnetic field is the gradual depopulation of the energetically higher spin-split Landau level.

We describe the phase diagram of our system using a model in the spirit of Ref.~\onlinecite{rosenow_influence_2007}, which incorporates similar ideas as Ref.~\onlinecite{evans_coulomb_1993}. We consider the system in thermodynamic equilibrium at a magnetic field $B_0$ and gate voltage $V_\mathrm{PG}^{(0)}$, where it minimizes its electrostatic energy by establishing a particular charge distribution in the \gls{QD}. When either the magnetic field changes by $\delta B$, or the gate voltage by $\delta V_\mathrm{PG}$, a charge imbalance will arise between the two compressible regions, which is given by
\begin{equation}
\begin{aligned}
\delta Q_1 &=  \Delta n_1 - \delta B \bar{A}/\phi_0 - C_1 \delta V_\mathrm{PG}/e\\
\delta Q_2 &=  \Delta n_2 + \delta B \bar{A}/\phi_0 - C_2\delta V_\mathrm{PG}/e.
\end{aligned}
\label{eq:imbalance}
\end{equation}
Here $\delta Q_i$ ($i=1,2$) describes the electron charge imbalance of the outer ($i=1$) and inner ($i=2$) compressible region in units of the elementary charge $e$, $\Delta n_i$ is the discrete change of charge in the respective region by single electron tunneling. The terms $\pm \delta B \bar{A}/\phi_0$ account for the flow of Hall current across the incompressible $\nu=1$ region upon a change in magnetic field, where $\bar{A}$ is the area enclosed by the incompressible region at $B_0$ and $\phi_0=h/e$ is the magnetic flux quantum. The $C_i>0$ are effective capacitances between the two compressible regions and the plunger gate. The system will react to a change in gate voltage $\delta V_\mathrm{PG}$ by increasing the charge imbalance in both compressible regions while simultaneously expanding the area enclosed by the incompressible region \cite{halperin_theory_2011}. The effective capacitances take these effects into account in leading order. The total capacitance between the plunger gate and the \gls{QD} is $C_\mathrm{PG}=C_1+C_2$. The change in the electrostatic energy of the system compared to the situation at $(B_0, V_\mathrm{PG}^{(0)})$ is in leading order given by
\begin{equation}
\delta E = \frac{1}{2}K_1\delta Q_1^2 + \frac{1}{2}K_2\delta Q_2^2 + K_{12}\delta Q_1\delta Q_2,
\label{eq:deltaE}
\end{equation}
where the $K_i$ are the charging energies of the two compressible regions, and $K_{12}$ describes the mutual capacitive coupling between them. In general, $K_1,K_2>K_{12}$. The system minimizes the energy functional \eqref{eq:deltaE} at given $\delta B$ and $\delta V_\mathrm{PG}$ by choosing appropriate values $\Delta n_1$ and $\Delta n_2$ in Eqs.~\eqref{eq:imbalance}. 

This model allows us to interpret the data in Fig.~\ref{fig2}(b) in terms of the indicated phase diagram. Charge rearrangements between the inner and the outer compressible region occur along horizontal white solid lines (\textcircled{\raisebox{-0.9pt}{3}}). Crossing a dashed white line (\textcircled{\raisebox{-0.9pt}{1}}) adds an electron to the outer compressible region, crossing a dotted white line (\textcircled{\raisebox{-0.9pt}{2}}) adds one to the inner compressible region. The hexagon-shaped regions of stable charge have a characteristic magnetic field period of one flux quantum per area $\bar{A}$ in agreement with Ref.~\onlinecite{mceuen_self-consistent_1992}, which allows us to extract $\bar{A}=\SI{0.48}{\micro m^2}$. This value is in good agreement with the zero magnetic field COMSOL simulations, if we calculate $\bar{A}$ as the area enclosed by the contour line of constant density $n_\mathrm{L}=eB_0/h$.
The characteristic gate voltage period of the phase diagram predicted by Eqs.~\eqref{eq:imbalance} is $e/C_\mathrm{PG}$. From the experimentally observed period of about $\SI{6}{mV}$ we extract $C_{PG}=\SI{27}{aF}$. Equations \eqref{eq:imbalance} predict lines of constant total electron number $N$ in the \gls{QD} to run exactly vertically, as approximately observed in Fig.~\ref{fig2}(b). They further predict lines of constant $\Delta n_1$ to have the (negative) slope $\delta B/\delta V_\mathrm{PG}=-C_1\phi_0/e\bar{A}$, whereas lines of constant $\Delta n_2$ have the (positive) slope
\begin{equation}
\left.\frac{\delta B}{\delta V_\mathrm{PG}}\right|_{\Delta n_2=\mbox{const.}}=\frac{C_2\phi_0}{e\bar{A}},
\label{eq:constN2}
\end{equation}
which allows us to estimate $C_1=\SI{16}{aF}$ and $C_2=\SI{11}{aF}$. With $K_1=\SI{220}{\micro eV}$ from the diamond measurement in Fig.~\ref{fig2}(c), we further determine the remaining model parameters $K_2=\SI{250}{\micro eV}$ and $K_{12}=\SI{143}{\micro eV}$.

On a larger magnetic field scale, the valleys of constant electron number $N$ in Fig.~\ref{fig2}(a) have a finite slope rather than being exactly vertical as predicted by the model. This observation implies a gradual redistribution of electrons between the quantum dot and the leads. Such an effect is due to the concerted action of the single-particle energies in dot and leads, which increase with magnetic field, and the capacitive coupling between dot and leads. The effect could be captured in a model that adds a term proportional to magnetic field and total electron number in the dot to the energy functional in Eq.~\eqref{eq:imbalance}, which we omit here for simplicity.

The measurements of the conductance as a function of magnetic field show a dependence on the previous history of the magnetic field sweeping. The general pattern of the Coulomb oscillations in Fig.~\ref{fig2}(a) is conserved while the exact peak positions shift. We assume that this form of hysteresis could be connected to dynamic nuclear spin polarization effects originating from the different spin polarization of the two spin branches of the lowest Landau level \cite{wald94}. Tunnelling between the two compressible regions is associated with a spin flip and changing the magnetic field monotonically causes only spin flips to one spin direction. We carefully checked that the measurements shown here are fully reproducible for sweeping the magnetic field in the same way. In this paper, we only present datasets that are comparable and measured with the same sweep direction.

\begin{figure}[htbp]
\includegraphics[width=\columnwidth]{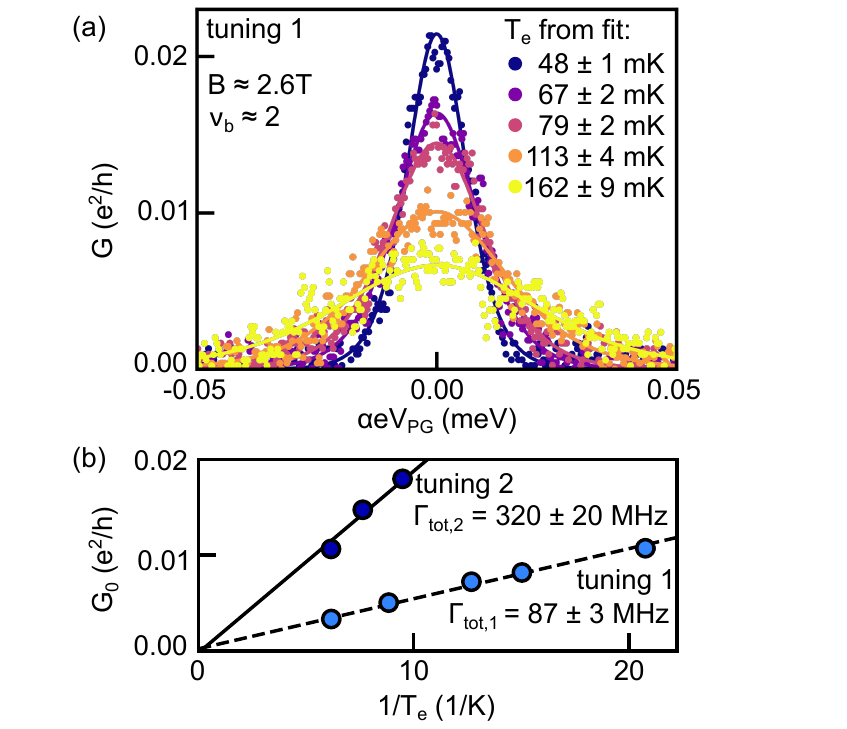}
\caption{(a) Conductance around filling factor $\nu_{\mathrm{b}} \approx 2$ as a function of the plunger gate voltage converted to energy measured for different electron temperatures $T_e$. Solid lines denote fits according to Eq.~(\ref{eq:SLT}). (b) Conductance peak height $G_0$ as a function of inverse electron temperature $1/T_e$ for two different coupling strengths of the \gls{QD} to the leads. The lines correspond to least square fits of the expected linear dependence yielding the effective dot--lead coupling $\Gamma_\mathrm{cl}$.
} 
\label{fig3}
\end{figure}

We now study the temperature-dependence of the conductance resonances in Fig.~\ref{fig2} to find out whether a single or multiple levels contribute.  Figure~\ref{fig3}(a) shows the temperature dependence of the conductance resonance measured along the blue arrow in Fig.~\ref{fig2}(b) at a bulk filling factor $\nu_{\mathrm{b}}\approx 2$. The magnetic field is carefully tuned around $B \approx 2.6~\mathrm{T}$ such that the Coulomb resonance is crossed in the middle between two charge rearrangements between inner and outer compressible regions.
The width of the resonance indicates thermal broadening. Fitting the measurements with the line shape valid for single-level transport \cite{averin_theory_1991,beenakker_theory_1991}
\begin{equation}
G = G_0 \cosh^{-2}\left[\dfrac{e\alpha_\mathrm{PG}(V_\mathrm{PG}-V_0)}{2k_\mathrm{B} T_\mathrm{e}}\right],
\label{eq:SLT}
\end{equation}
where $G_0 = e^2\Gamma_\mathrm{cl}/4k_\mathrm{B} T_\mathrm{e}$, with fitting parameters $V_0$, $G_0$, and $T_\mathrm{e}$, leads to the temperatures indicated in the figure legend. Here, $V_0$ is the resonance position in gate voltage, $\alpha_\mathrm{PG}$ the plunger-gate lever arm, $T_\mathrm{e}$ is the electron temperature, and $\Gamma_\mathrm{cl}$ is the (temperature-independent) classical tunneling rate through the dot. Surprisingly, increasing the electron temperature reduces the peak amplitude $G_0$ of the conductance resonance. Figure~\ref{fig3}(b) shows for two different values of $\Gamma_\mathrm{cl}$ that $G_0$ is a linear function of $1/T_\mathrm{e}$, as expected for single-level transport. For multi-level transport, one would expect $G_0$ to be independent of temperature as more and more levels contribute to the conductance resonance, thereby cancelling the $1/T_e$ dependence of the individual levels \cite{beenakker_theory_1991}.

Taking the previously determined $\bar{A}$ as the approximate size of the \gls{QD}, we estimate the order of magnitude of the zero magnetic field excitation energy to be $\Delta = \pi\hbar^2/m^\star\bar{A}\approx \SI{7.4}{\micro eV}$, which is comparable or even smaller than the thermal smearing of about $3.55k_\mathrm{B}T_\mathrm{e}$ in our experiment. This energy spacing is not compatible with the observed single-level transport. Alternatively, in agreement with Ref.~\onlinecite{chamon97}, we estimate the typical energy spacing $\Delta$ of edge excitations in a single-particle picture using the flux quantization rule $B\Delta A=\phi_0$, where $\Delta A$ is the difference of area enclosed by two edge excitations neighboring in energy. Assuming a circular \gls{QD} with the steepness $V'$ of the edge potential leads to $\Delta_0 = hV'/eBL=hv_0/L$, with $v_0$ being the drift velocity of edge excitations and $L$ the perimeter of the \gls{QD}. While our derivation of this result uses the single-particle picture, which involves the steepness $V'$ of a Hartree-potential, the final result expressing $\Delta_0$ in terms of a  velocity $v_0$ does also hold in a Luttinger-liquid description \cite{Wen90} when the drift velocity is replaced by the full velocity $v$, which includes interaction effects. In such a description, $v$ has the meaning of the propagation velocity of edge magnetoplasmon excitations arising under the assumption of short-range intra-edge Coulomb interactions, and is given by $v = v_0 + v_\mathrm{C}$ with $v_\mathrm{C}=(e^2/4 \pi^2 \hbar \epsilon \epsilon_0)\ln(d/\ell_\mathrm{c})$. Here, $\epsilon$ denotes the relative dielectric constant of the host material, $d$ the distance to a screening gate, and $\ell_\mathrm{c}=\sqrt{\hbar/eB}$ the magnetic length. For $d/\ell_\mathrm{c} = 10$ one finds $v_\mathrm{C} = \SI{1.23e5}{m/s}$ in GaAs. Using Luttinger liquid theory to compute the electrostatic charging energy $K_1$ from the zero mode energy, one finds that $K_1 \equiv \Delta = h v/L $. Using the value for $K_1$ discussed above, one can estimate the level spacing to be of the order of $\SI{220}{\micro eV}$, much larger than the thermal energy, in agreement with the observed single-level transport. This relation also provides a value for the velocity $v = \SI{1.33e5}{m/s}$ (using $L = 2 \sqrt{\pi A} = \SI{2.5}{\micro m}$ for a circular QD with area $A$), somewhat larger than the Coulomb velocity $v_C$ discussed above. Alternatively, taking the highest measured electronic temperature of $\SI{162}{mK}$ as a lower limit for $\Delta$, we obtain $\Delta > \SI{50}{\micro eV}$, and thereby $v > \SI{0.29e5}{m/s}$. We see that the edge excitation energy of a large QD in the quantum Hall regime can exceed the zero-field excitation energy by almost an order of magnitude.

In Ref.~\onlinecite{mcclure_09}, the relation $\Delta=hv/L$ was used to estimate the velocity of edge-excitations in a \gls{QD}, giving $v\approx \SI{1e5}{m/s}$ for a magnetic field of $\SI{1}{T}$ which is comparable with our value based on the Luttinger theory. In Ref.~\onlinecite{nakamura_aharonovbohm_2019}, a velocity between $v=\SI{4e4}{m/s}$ and $\SI{8e4}{m/s}$ around $B=\SI{2}{T}$ was found, which is also comparable with our lower bound discussed above. 

\section{Evolution to filling factor $\mathbf{\nu_{\mathrm{b}} = 1}$}
\begin{figure}[htbp]
\includegraphics[width=\columnwidth]{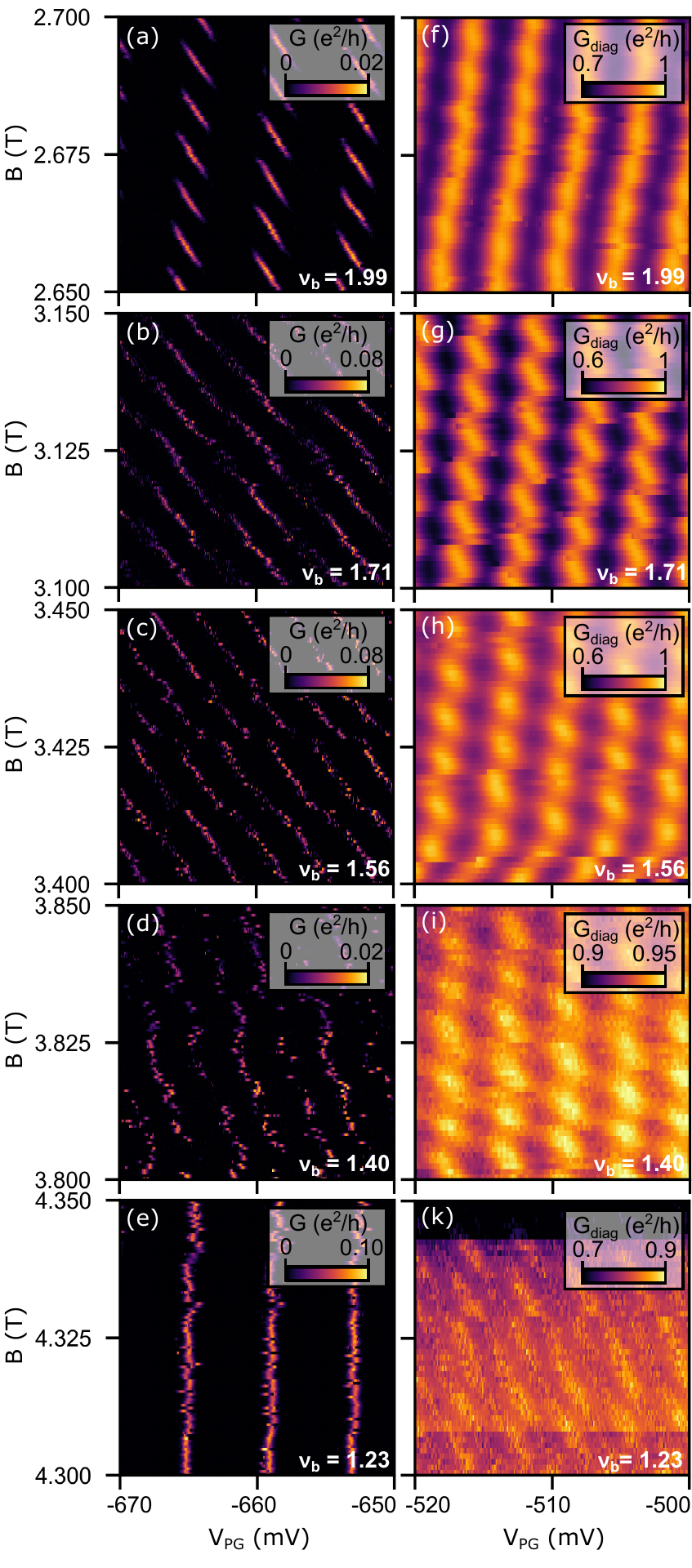}
\caption{(a)~-~(e) \Gls{QD} operated in the strong Coulomb blockade regime. Conductance $G$ through the \gls{QD} plotted as a function of plunger gate voltage and magnetic field for different magnetic field ranges between filling factor $2>\nu_{\mathrm{b}}>1$. (f)~-~(k) \Gls{QD} operated as an interferometer with the barriers transmitting one edge channel ($t_{\mathrm{barrier}} \approx 0.9$). Conductance $G_\mathrm{diag}$ measured from the transverse voltage across the \gls{QD} ($V_{\mathrm{diag}}$, see Fig.~\ref{fig1}~(a)) shown as a function of plunger gate voltage and magnetic field for the same magnetic field ranges as in (a)~-~(e).} 
\label{fig5}
\end{figure}

With the present understanding of Fig.~\ref{fig2} in mind, we now proceed with investigating the evolution of the pattern of conductance resonances with decreasing filling factor in the range $2\gtrsim\nu_{\mathrm{b}}\gtrsim 1$.
Generally, we observe that increasing the magnetic field in this range strongly reduces the tunnel coupling of the \gls{QD} to the leads. In order to keep approximately the same coupling, the gates LB and RB have to be retuned at each new magnetic field.
Figures~\ref{fig5}~(a--e) show the conductance as a function of plunger gate voltage and magnetic field at different filling factors in this range. The different color scales used for the different filling factors originate from retuning the tunnel barriers.
Figure~\ref{fig5}(a) reproduces Fig.~\ref{fig2}(b).
In Fig.~\ref{fig5}(b,c), decreasing the filling factor towards $\nu_{\mathrm{b}} = 1$, the magnetic field period of the resonances increases [particularly visible comparing (b) and (c)]. This observation indicates a contraction of the incompressible $\nu=1$ region and a concomitant reduction of the area $\bar{A}$ of the inner compressible region with increasing $B$.

At the same time, the visible charging lines extend into the regions of constant total electron number $N$. While vertical regions of suppressed conductance corresponding to constant $N$ are still discernible in Fig.~\ref{fig5}(b), the lines extend so strongly in (c), that the number of resonances crossed in the image at constant $B$ doubles as compared to (a). It has been experimentally observed \cite{van_der_vaart_time-resolved_1994,van_der_vaart_time-resolved_1997} that the tunneling rate between the inner and outer compressible regions reduces dramatically with decreasing filling factor. Eventually, the processes become so slow that individual tunneling events occur on time scales of seconds, which can be measured in real time. An increase of the width of the incompressible $\nu=1$ region with $B$ also predicted theoretically \cite{lier_gerhards_94} accounts for this observation. We interpret the conductance resonances extending into the regions of constant electron number $N$ as a result of the competition between the tunneling rate between inner and outer compressible region, which tends to return the system to thermodynamic equilibrium after changes of $V_\mathrm{PG}$, and the time scales of the measurements. Slow relaxation to equilibrium extends the observed charging lines into the hexagonal regions of the phase diagram, where they would be absent in thermodynamic equilibrium in the dot. This interpretation is corroborated by the time-resolved observation of electron tunneling in our experiment (not shown), and by the increased 'noisiness' of the data with increasing magnetic field, observable in Figs.~\ref{fig5}(a--d).

The slope $\delta B/\delta V_\mathrm{PG}$ of the observed resonances in Fig.~\ref{fig5}(a--e) is negative at all filling factors. Its absolute value, however, decreases from (a) to (b) and then increases continuously from (b) to (e). Within our model described by Eqs.~\eqref{eq:imbalance} and \eqref{eq:deltaE}, we find
\begin{equation}
\left.\frac{\delta B}{\delta V_\mathrm{PG}}\right|_\mathrm{res} = -\frac{(K_1C_1+K_{12}C_2) \phi_0}{e\bar{A}(K_1-K_{12})}.
\label{eq:resslope}
\end{equation}
The detected decreasing area $\bar{A}$ will tend to increase the magnitude of the slope as observed in Fig.~\ref{fig5}(b-e). We speculate that the initial decrease between (a) and (b) is due to the interaction parameters $K_1$, $K_{12}$, $C_1$ and $C_2$, for which a quantitative theory is not available.

In the limit of filling factor $\nu=1$ in the quantum dot, we expect $\bar{A}\rightarrow 0$, leading within our model to an infinite magnetic field period, and an infinite negative slope of the resonances, as observed in Fig.~\ref{fig5}(e). The observation leads to the conclusion that the filling factor in the dot $\nu\approx 1$ has been reached, despite the somewhat higher bulk filling factor $\nu_\mathrm{b}=1.23$. The relatively strong density reduction in the dot compared to Fig.~\ref{fig1}(b) is the result of the strong negative voltages on the gates RB and LB [c.f. Fig.~\ref{fig1}(a)]. The spacing of the resonances in gate voltage that was given by $e/C_\mathrm{PG}$ close to $\nu_{\mathrm{b}}=2$ is about the same at $\nu_{\mathrm{b}}=1$. This is due to the fact that the decrease in $C_2$ with increasing magnetic field is compensated by an increase of $C_1$, such that their sum $C_\mathrm{PG}$ stays roughly constant.

\begin{figure}[htbp]
\includegraphics[width=\columnwidth]{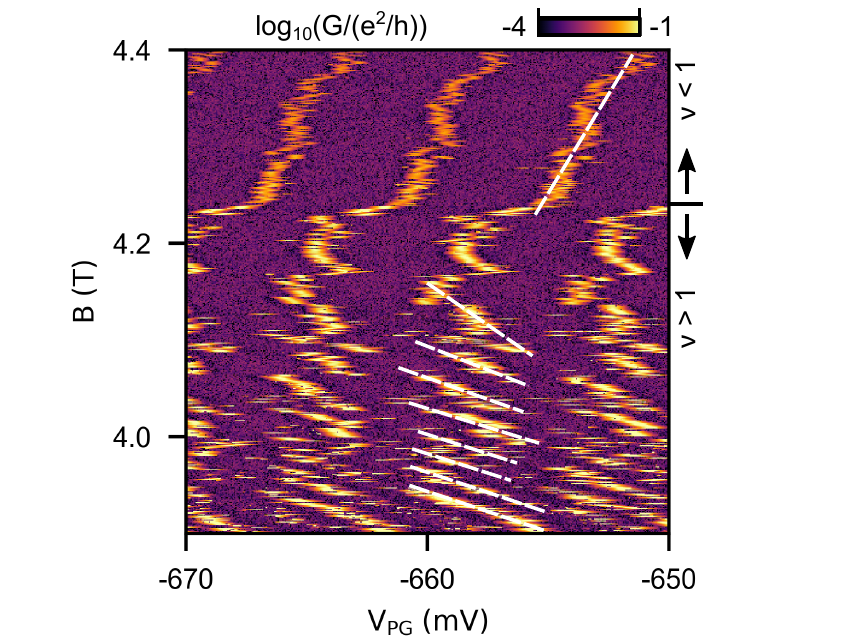}
\caption{Logarithm of the conductance through the QD as a function of plunger gate voltage and magnetic field. Dashed lines indicate the trend of the resonances. Around $\SI{4.22}{T}$ the upper Landau level becomes completely depopulated.} 
\label{fig6}
\end{figure}

The depopulation of the inner Landau level can be seen very clearly by looking at the resonances on an extended magnetic field scale as it is shown in Fig.~\ref{fig6}. At magnetic fields below \SI{4.22}{T} the Coulomb resonances show a pattern reminiscent of the phase-diagram near $\nu_{\mathrm{b}}=2$ indicating the presence of charge transfer from the upper to the lower Landau level. Beyond $B = \SI{4.22}{T}$ the behavior changes abruptly, presumably because the upper Landau level is essentially depopulated.

\section{Interferometer}
\begin{figure*}[t]
\includegraphics[width=\textwidth]{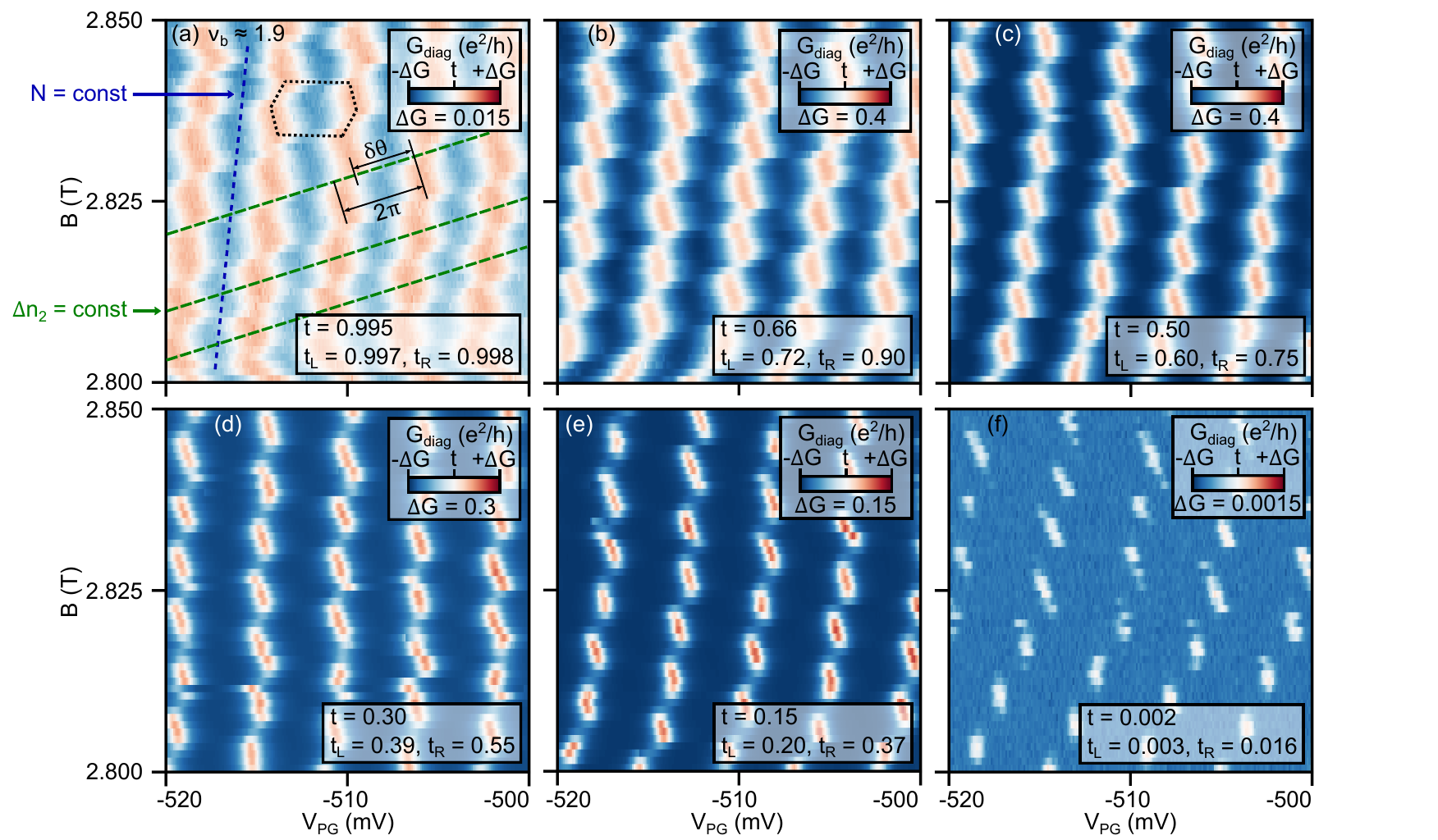}
\caption{(a-f) QD operation continuously changed from  interferometer to strongly Coulomb blockaded QD. Conductance $G_{\mathrm{diag}}$ shown as a function of plunger gate voltage and magnetic field for different, decreasing transmission of the left and right barrier ($t_\mathrm{L}$, $t_\mathrm{R}$) corresponding to set points of the barrier voltages $V_\mathrm{LB}$, $V_\mathrm{RB}$ indicated in Fig.~\ref{fig1}(c-e). The classical transmission value $t$ is calculated according to Eq.~(\ref{classical_transmission}) and the color scale is chosen around the classical value $t$ (white) for each plot.}
\label{fig7}
\end{figure*}

We now change the transmission of the \glspl{QPC} that form the barriers of the \gls{QD} such that we almost completely transmit one of the edge channels inducing only weak backscattering [Fig.~\ref{fig1}(c)]. Thereby we operate the device as a quantum Hall interferometer \cite{chamon97,halperin_theory_2011, ji_electronic_2003, zhang_distinct_2009, nakamura_aharonovbohm_2019}. This is achieved using $V_\mathrm{LB}=\SI{-0.86}{V}$ and $V_\mathrm{RB}=\SI{-1.43}{V}$ [c.f. Fig.~\ref{fig1}(c)]. The transmissions of the outermost edge channel of the left and right \gls{QPC} are $t_L = 0.997$ and $t_R = 0.998$ respectively, giving the incoherent transmission $t=0.995$ according to
\begin{equation}
t = \frac{t_L t_R}{1-(1-t_L)(1-t_R)}.
\label{classical_transmission}
\end{equation}
Since the device resistance is relatively small in this regime, of the order of the resistance quantum $h/e^2$, we switch to a four-terminal measurement of $G_\mathrm{diag}$.

The diagonal conductance is shown in Fig.~\ref{fig7}(a) for a filling factor $\nu_{\mathrm{b}} \approx 1.9$. 
We choose the color scale in Fig.~\ref{fig7}~(a) such that the classical transmission value appears white, while positive (negative) deviations from this value are shown in red (blue). The observed conductance shows a periodic pattern around the classically expected value.
Despite the strong coupling of the outer edge channel to the leads, the pattern shows striking similarities to the phase diagram in Fig.~\ref{fig2}(b) measured in the weak dot--lead coupling limit. Resonances appearing in red are reminiscent of the tunneling resonances into the outer compressible region of the lower Landau level. They appear to be strongly broadened and their modulation amplitude is tiny as compared to Fig.~\ref{fig2}(b) as a result of the strong tunnel coupling.
Their period in gate voltage is slightly reduced to $\SI{4.4}{mV}$ and the magnetic field period of the pattern is $\SI{7.1}{mT}$ giving an area $\bar{A}=\SI{0.58}{\micro m^2}$, slightly larger than the area estimated from Fig.~\ref{fig2}(b). Both changes are compatible with the larger dot area caused by opening the tunnel barriers. The similarity between the observed conductance pattern in the interferometer and in the Coulomb blockade is one of the main findings of this paper. It agrees qualitatively with the findings in Ref.~\onlinecite{Sivan2016} and the theoretical prediction in Ref.~\onlinecite{Stern2010}.

As a consequence, a phase diagram similar to that in Fig.~\ref{fig2}(b) can be constructed in the open regime, as indicated by the hexagon outlined with a black dotted line in Fig.~\ref{fig7}(a).
If we follow lines of a constant electron number $\Delta n_2$ in the phase diagram [green dashed lines in Fig.~\ref{fig7}(a)], the observed conductance oscillations in Fig.~\ref{fig7}(a) represent the Fabry--P\'{e}rot resonances of the interferometer. Although the lower Landau level couples strongly to the leads in this regime, the upper Landau level remains very weakly coupled to the lead and to the outer compressible region due to the presence of the incompressible $\nu=1$ region, which completely surrounds it. In contrast to the case of the \gls{QD} in the weak coupling limit, this incompressible region is now no longer confined inside the \gls{QD}, but it now extends through the constriction into the lead regions. Abrupt phase jumps of the interference pattern arise, whenever the number $\Delta n_2$ changes due to the electron-by-electron depopulation of the upper Landau level with increasing magnetic field. These phase jumps have been theoretically described in Ref.~\onlinecite{halperin_theory_2011} (see also below).

The full evolution from the regime, where the outer compressible region of the \gls{QD} is strongly coupled to the leads to the regime of weak coupling with strong charge quantization in the outer compressible region is shown in the sequence of measurements in Fig.~\ref{fig7}(a--f). Along this sequence, the classical transmission $t$ is tuned following the blue points in Fig.~\ref{fig1}(c) from $t=0.995$ to $t=0.002$, i.e., from the Fabry--P\'{e}rot quantum Hall interferometer regime to the Coulomb blockade regime. At each stage of tuning, the hexagonal phase diagram is discernible indicating that the physics of the fully Coulomb blockaded \gls{QD} in the quantum Hall regime carries over to the interferometer regime. Similar observations were made in Ref.~\onlinecite{ofek_role_2010} in larger interferometers, in which the phase jumps were not observed.

In the following, we complement the experimental investigation of the transition from Coulomb blockaded \gls{QD} to interferometer by a comparison to a theoretical model.
Within a slight extension of the existing theory of the Fabry--P\'{e}rot quantum Hall interferometer of Ref.~\onlinecite{halperin_theory_2011}, Eqs.~\eqref{eq:imbalance} are replaced by
\begin{equation}
\begin{aligned}
\delta Q_1 &= \frac{B_0(A-\bar{A})}{\phi_0} - \delta B \bar{A}/\phi_0 - C_1\delta V_\mathrm{PG}/e \\
\delta Q_2 &= \Delta n_2 + \delta B \bar{A}/\phi_0 - C_2 \delta V_\mathrm{PG}/e.
\end{aligned}
\label{eq:interferometer}
\end{equation}
The previous change in electron number $\Delta n_1$ of electrons in the outer compressible region of the \gls{QD} is in the interferometer case no longer bound to be integer. The system can react to a change in magnetic field or gate voltage by changing the interferometer area by $A-\bar{A}$, which is a continuous variable. However, the inner incompressible region still has discrete charge, and $\Delta n_2$ is required to be integer. The system minimizes the energy functional \eqref{eq:deltaE} at given $\delta B$ and $\delta V_\mathrm{PG}$ by choosing appropriate values for $\Delta n_2$ and $(A-\bar{A})$.

According to this model, lines of constant $\Delta n_2$ have a slope $\delta B/\delta V_\mathrm{PG}$ exactly given by Eq.~\eqref{eq:constN2}, i.e., the same as in the case of Coulomb blockade [green dashed lines in Fig.~\ref{fig7}(a)]. The period of resonances as a function of $\delta V_\mathrm{PG}$ at constant magnetic field is found by considering the change of phase of the interfering electrons upon a change of area, given by \cite{halperin_theory_2011}
\begin{equation}
\delta\theta = 2\pi\frac{B_0(A-\bar{A})}{\phi_0}. 
\end{equation}
Going from one resonance to the next along the green dashed lines winds the phase by $2\pi$ [as indicated in Fig.~\ref{fig7}(a)], which can be caused by a gate-voltage increase of $e/C_\mathrm{PG}$ according to Eqs.~\eqref{eq:interferometer}. This means that the vertically aligned resonances in Fig.~\ref{fig7}(a) correspond to the Fabry--P\'{e}rot resonances. They originate from resonances in the Coulomb blockaded \gls{QD}, where the total electron number $N$ in the dot changes by one, as is experimentally evident from the evolution in Figs.~\ref{fig7}(a--f). These resonances represent lines of constant interferometer phase (area), and their negative Aharonov--Bohm-like slope is given by Eq.~\eqref{eq:resslope}, like in the \gls{QD} limit.

Furthermore, the model predicts an electron-by-electron depopulation of the upper Landau level with increasing magnetic field, which is responsible for the phase jumps oberserved with a period of $\phi_0/\bar{A}$ as the magnetic field is increased. As long as $\Delta n_2$ and $\delta V_\mathrm{PG}$ are constant, the area of the interferometer will increase with increasing magnetic field. Once a single electron redistributes between the upper and the lower Landau level, the interferometer area jumps by $\Delta A=(K_{12}/K_1)(\phi_0/B_0)$. Thereby, on a larger magnetic field scale, the interferometer area oscillates around a constant value with a magnetic field period of one flux quantum. The discrete jumps in interferometer area correspond to discrete jumps $\Delta\theta=2\pi K_{12}/K_1$ of the phase difference between the interfering electron waves. This phase jump can be directly read from the measurement in Fig.~\ref{fig7}~(a) as indicated. The fact that this phase jump is larger than $\pi$ indicates that the interferometer is in the so-called Coulomb-dominated regime \cite{halperin_theory_2011}.

The vertical alignment of the resonances corresponding to constant interferometer phase is in agreement with previous experiments on larger interferometers, in which the phase jumps were not observed, but vertical Fabry-P\'{e}rot resonances were seen \cite{ofek_role_2010}. In our experiment, the observation of the phase jumps enables the detection of the one-by-one redistribution of quasiparticles from the inner compressible region into the outer interfering edge channel.

The evolution of the interferometer transmission when changing the filling factor from $2>\nu_{\mathrm{b}}>1$ is shown in Fig.~\ref{fig5}(f-k).
Like in the quantum dot case, the voltages applied to the gates LB and RB had to be readjusted, when the filling factor $\nu_{\mathrm{b}}$ was changed.  The measurements in this figure can be readily interpreted in terms of the observations and the model discussed before. At all filling factors, except at $\nu_{\mathrm{b}}=1.23$ in Fig.~\ref{fig5}(k), the vertical Fabry--P\'{e}rot resonances are clearly discernible. At $\nu_{\mathrm{b}}=1.99$ in Fig.~\ref{fig5}(f), the phase jumps in $B$ are strongly washed out, indicating relatively strong tunnel coupling between the inner and the outer compressible regions. This coupling is weaker in Fig.~\ref{fig5}(g) at $\nu_{\mathrm{b}}=1.71$, where the phase jumps appear to be sharper. Figure~\ref{fig5}(h) is particularly interesting, because  the $\phi_0$-period in $B$ is visible even within the Fabry--P\'{e}rot minima. At the same time, the phase jumps along Fabry--P\'{e}rot resonances are strongly washed out, which leads to a superior visibility of these resonances. While this remains true in Fig.~\ref{fig5}(i), the visibility of the Fabry--P\'{e}rot resonances has become somewhat weaker, most likely for the same reason invoked for explaining the extending Coulomb resonances in Fig.~\ref{fig5}(c), i.e., the tunneling time competing with the measurement time. This effect appears to be very strong in Fig.~\ref{fig5}(k), where the upper Landau level is close to being completely depopulated. The fact that the charge redistributions between the two Landau levels is 'quenched' on intermediate time scales  allows for the observation of Aharonov--Bohm type of interference in this otherwise Coulomb-dominated interferometer.

\section{Conclusion}
In this paper, we have shown transport measurements of a large quantum dot containing about 500 electrons in the quantum Hall regime for filling factors $2 > \nu > 1$. We have experimentally investigated both the transition from an open Fabry--P\'{e}rot quantum Hall interferometer to a closed Coulomb blockaded quantum dot, as well as the evolution of transport from $\nu=2$ to $\nu=1$. We have compared our experimental results with theoretical models in the Coulomb blockaded and the interferometer regimes and found these models to be in agreement with the experimental observations, confirming the validity of the physical ideas presented in recent theoretical works \cite{rosenow_influence_2007,halperin_theory_2011}. A quantitative analysis allowed us to extract all relevant model parameters from the measurements. Model and experiment lead to the interpretation in terms of self-consistently forming compressible and incompressible regions within the quantum dot. The interplay between the charge quantization in the quantum dot and the electron-by-electron depopulation of the higher Landau level leads to a hexagonal phase diagram in the gate-voltage magnetic field plane, that bears similarity to the charge stability diagram of a double-quantum dot at zero magnetic field. In agreement with previous authors \cite{van_der_vaart_time-resolved_1994,van_der_vaart_time-resolved_1997}, we find a strong reduction of the tunnel coupling between the two compressible regions relating to the two participating Landau levels with increasing magnetic field. Temperature-dependent measurements in the Coulomb blockade regime allowed us to estimate the energy spacing of edge-excitations in the lower Landau level, which is significantly larger than the single-particle level spacing of the \gls{QD} at zero magnetic field. The close similarity between interferometer data and data in the Coulomb blockade regime show that the interferometer inherits the properties of the Coulomb blockaded quantum dot non-withstanding its strong coupling to the leads. The observed phase jumps of the Fabry-P\'{e}rot resonances in the interferometer, which are due to the discrete depopulation of quasiparticles from the inner compressible region of the quantum dot, offer the opportunity to detect quasiparticle tunneling and the concomitant charge redistribution. We anticipate that our results clarify the close relationship between Coulomb blockaded quantum dots and interferometers in the quantum Hall regime, and thereby enable the design of novel experiments that extend into the fractional quantum Hall regime in the future.
%

\begin{acknowledgments}
The authors acknowledge the support of the ETH FIRST laboratory and the financial support of the Swiss Science Foundation (Schweizerischer Nationalfonds, NCCR QSIT). B.~R. would like to acknowledge support by DFG grant RO 2247/11-1.
\end{acknowledgments}

%

\end{document}